# Enhancing sensitivity of atomic microwave receiver based on a laser array

Bo Wu [1], Ruiqi Mao [1], Yi Liu [1], Di Sang [1], Yanli Zhou [2], Yi Lin [1], Qiang An [1]*, and Yunqi Fu [1]*

[1] College of Electronic Science and Technology, National University of Defense Technology, Changsha 410073, China

[2] Institute for Quantum Science and Technology, College of Science, National University of Defense Technology, Changsha 410073, China

**Abstract**: Rydberg atom, which exhibits a strong response to weak electric fields, is regarded as a promising atomic receiver to surpass sensitivity of conventional receivers. However, its sensitivity is strongly limited by the noise coming from both classical and quantum levels and how to enhance it significantly remains challenging. In our experiment, 4.6 dB *SNR* enhancement is achieved by utilizing the 2 × 2 probe beam array, compared to the performance of a laser beam, and it can be enhanced further just by adding more laser beams. In particular, an experimental verification is given to 2 × 2 laser array is 19 nV/cm/Hz$^{1/2}$@8.57 GHz. More importantly, the probe laser array has coherence. Therefore, the SNR can still be improved when detecting the weak target signal in strong noise and clutter environment. The results could offer an avenue for the design and optimization of ultrahigh-sensitivity Rydberg atomic receivers and promote applications in cosmology, meteorology, communication, and microwave quantum technology.

# Introduction

Precision measurement technology based on Rydberg atomic receiver with high sensitivity, [1-3] broadband response characteristics, [4] small system size [5] and concealed anti-damage detection, [6-8] has great potential for applications in a diversity of fields, such as communication signals, [9-15] stereo reception, [16] Rydberg microwave-frequency-comb spectrometer, [17] imaging [18] and so on. The most exciting thing is that its theoretical sensitivity limit is -220 dBm/Hz, [19] which surpasses the classical receiver's sensitivity limit of -174 dBm/Hz.

So far, no experiment confirms that the atomic receiver sensitivity exceeds the traditional receiver sensitivity. Therefore, it remains elusive in the how to improve further sensitivity. Up to now there have existed several limits including classical noise (CN), quantum photon shot noise (QPSN) and quantum projection noise (QPN), which are the obstacles to further improving the atomic receiver sensitivity. To reach high sensitivity in the atomic receiver, the CN, QPSN and QPN are usually minimized as low as possible. Previous work has suggested to utilize muti-photon, [20] via homodyne readout to reduce photon shot noise with 300 μV/cm/Hz$^{1/2}$ electric(E)-field measurement sensitivity. Other experiments explore three-photon excitation to eliminate Doppler-broadening. [21] In microwave domain, experiments explore the superheterodyne systems [22-24] and E-field local enhancement. [25-31] Superheterodyne systems' sensitivity can be improved by optimizing the local oscillator intensity and intermediate frequency to maximize responsivity with 49 nV/cm/Hz$^{1/2}$ .[22] Typical examples of E-field local enhancement are split-ring resonator which is utilized at 1.3 GHz to achieve a 100 times E-field enhancement with 5.5 μV/cm/Hz$^{1/2}$ . [29] This method is useful at lower frequencies, but is difficult to scale to shorter microwave wavelengths. Moreover, further studies have been performed using many-body Rydberg atomic system. [32] But phase transition process of many-body system is difficult to build. In addition, since the atomic receiver sensitivity is limited by the quantum projection noise limit ( $S \propto 1/\sqrt{N}$ , where $N$ is the Rydberg atoms population), [19] the sensitivity can be enhanced by repump (two times enhancement), [33] increasing vapor cell length [34,35], as illustrated in Fig. 1(a). However, two limitations prevent it from being utilized in an overlong vapor cell to further improve sensitivity. To begin with, the laser intensity drops when the transmission distance rises. It contributes to a large difference in probe and coupling laser intensity between the two ends of the cell. This ultimately causes a poor electromagnetically induced transparency (EIT) spectrum. Furthermore, an overlong vapor cell does not maintain the average E-field amplitude constant throughout cell lengths because the high-order effect cannot be eliminated by MW E-field inhomogeneity. [36] Therefore, we can increase the probe laser waist and the probe laser array approach to excite more Rydberg atoms population. But increasing the probe laser waist will lead to a decrease in instantaneous bandwidth. [37] Finally, a laser array method becomes a promising candidate to further enhance the sensitivity in this scenario. We combine the concept of antenna array in microwaves, and multiple beams are equivalent to multiple Rydberg receivers. Additionally, the maximum laser output power is 300.0 mW, as demonstrated the DL Pro. However, the optimal probe laser power is generally 0.1 mW. This mismatch leads to low utilization of the DL Pro's potential capabilities. Therefore, the laser array method can ensure the full utilization of the laser's output power.

In this work, the laser array method is reported for the first time to further improve sensitivity. To start with, we utilize more beams (a 2 × 2 probe laser array) to improve the *SNR* of the atomic receiver. Split-beam grating and collimator grating are employed to divide a probe laser beam into four parallel uniform beams. The 2 × 2 probe laser array both passes through the same vapor cell for detection, which has much better consistency than the laser array illuminated respectively multiple vapor cells. We focus on influences of the 2 × 2 probe laser array in the EIT signal height, the beat-note frequency probe laser transmission signals time domain waveform, the noise spectrum and *SNR* enhancement with different beam numbers. Furthermore, we combine the sensitivity enhancement methods including microwave LO and microwave E-field local enhancement (utilizing a copper reflective plate to constitutive a standing wave system) to demonstrate the laser array technique can significantly multiply the sensitivity when the microwave domain enhancement approach works at its upper limit. Finally, we provide a sensitivity analysis approach that is appropriate for comparing two fundamentally different types of receivers: conventional and atomic. Therefore, in this paper, we reveal the importance of the laser array in practical applications and shed new light on the guidance for the design and optimization of ultrahigh-sensitivity Rydberg atomic receivers.

## Methods

Two receiver sketches are described comparing an atomic microwave receiver with conventional receivers, as depicted in Fig. 1(b, c). The vapor cell has cesium atoms that are driven to the Rydberg state by the probe and coupling lasers. The transmission signal of the probe laser can be detected on a photodetector (PD). To resolve the phase, a coupler and a microwave local oscillator (LO) can be used. Conventional receivers require antennas to receive electromagnetic waves and convert them into high-frequency currents. They need filters, low-noise amplifiers (LNA), mixers, and other components for signal processing.

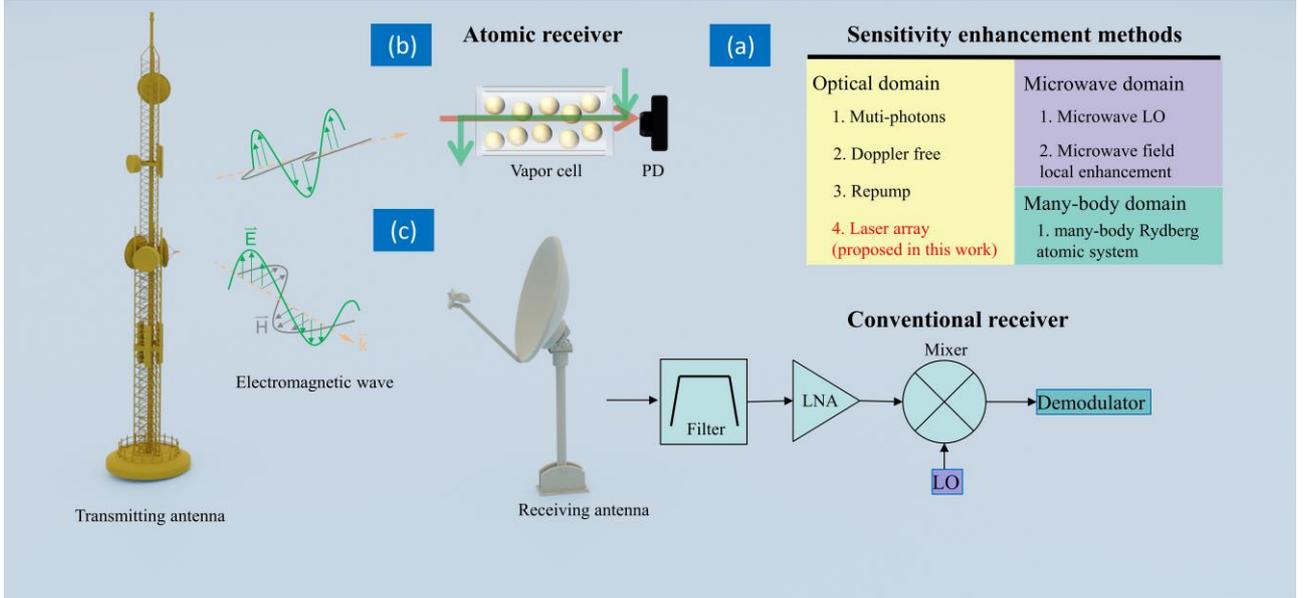

Fig. 1 | (a) Optical domain methods to enhance sensitivity, including using multiple photons, doppler free, an additional repump laser and a laser array (proposed in this work). Microwave domain methods include adding a microwave LO and E-field local enhancement. Many-body domain method include many-body Rydberg atomic system. Sketch of basic Rydberg-atom-based microwave receiver (b) and conventional receiver (c).

**Measurement method of microwave electric field sensitivity**

The AT splitting is proportional to the E-field and is perceived as: [9,38]

$$E = \frac{\hbar}{\mu}\Omega_{\mathrm{MW}} = 2\pi\frac{h}{\mu}\Delta_{\mathrm{AT}}, \tag{3}$$

where $\Omega_{\mathrm{MW}}$ is the Rabi frequency of the Rydberg state transition, $\mu$ is the transition dipole moment of the adjacent Rydberg states, $\hbar$ is reduced Planck's constant, and $\Delta_{\mathrm{AT}}$ is AT split peak spacing.

When the SIG E-fields is small and the AT splitting interval is close to or less than the EIT spectral linewidth, the method of measuring the splitting interval directly by reading the spectral maxima fails to measure it accurately. [8] Subsequently, we use the Rydberg atomic superheterodyne receiver to continuously reduce the SIG power $P$ of the horn source until the SA does not detects the beat signal, at which point the SIG source output power can be substituted into the antenna radiation formula to calculate the E-fields (sensitivity) reaching the vapor cell. The antenna radiation formula can be expressed as (Supplementary Information S1 for the derivation)

$$E = \mathcal{F}\frac{1}{\sqrt{2\pi c\varepsilon_0}}\frac{\sqrt{PGL}}{R} = k_{\mathrm{a}}\sqrt{P}, \tag{4}$$

where $c$ is the light speed in vacuum, $\varepsilon_0$ is the permittivity of free space, $R$ is the distance from antenna to the laser beam,

$\mathcal{F}$ is the perturbation factor caused by space scattering and standing wave (or resonances) disturbance in the vapor cell [39, 40], $PGL$ is the radiated power of MW ($PGL = P + G - L$, $P$ represents the output power of signal source, $G$ represents the gain of antenna, $L$ represents the insertion loss of transmission line). In the experiment, all the above parameters are constant, except for $P$, which means $k_a$ is constant. Coefficient $k_a = \mathcal{F} \frac{\sqrt{GL}}{\sqrt{2\pi c \varepsilon_0 R}}$ can be calculated by measuring the typical EIT-AT splitting spectral lines. Consequently, the SIG power $P$ can be directly converted to the E-fields amplitudes.

## Results and discussions

### Experiment

Fig. 2(a) illustrates the experimental setup overview. The red frame is the 852 nm laser frequency locking module, the green frame is the 509 nm laser frequency locking module, and the blue frame is the experimental measurement module. A strong LO MW E-fields drive a transition between two different Rydberg states $44D_{5/2}$ and $45P_{3/2}$. The strong LO MW and weak SIG MW fields are generated via two signal generators. Then, the two signals are synthesized by a 2-way microwave resistive power divider. Finally, the synthesized signal is radiated by a one horn antenna. The far field distance is 18 cm for the horn antenna with 44 mm × 34 mm physical size. Hence, we positioned the antenna 25 cm away from the vapor cell. In the laser overlap area, there is a cylinder-shaped cesium vapor cell that is 50 mm long and 10 mm in diameter at room temperature. In the vapor cell, the probe and the coupling laser are counter-propagated and overlapped to constitute the EIT process. After a PD detects the probe laser, it is connected to a spectrum analyzer. Superheterodyne sensitivity is theoretically defined as the minimum detectable power when the $SNR$ decreases to 1. In our demonstration, the LO MW signal frequency is 8.57 GHz, while the weak SIG signal frequency is set to 8.57 GHz + $\delta_s$ kHz. Subsequently, once the coupling laser is locked, the probe laser intensity will oscillate at a $\delta_s$ beat-note frequency. The beat-note signal amplitude reflects the ability to detect a weak E-fields. The spectrum analyzer measures PD output signal voltage with a 1 Hz resolution bandwidth (one second averaging time).

As shown in Fig. 2(b1, b2), microscopic images of gratings are displayed. A split-beam grating divides a laser beam into four divergent laser beams. The 2×2 probe laser array is collimated utilizing a collimator grating, with a 1.6 mm spacing between adjacent spots. Fig. 2(c1, c2) show transverse profiles of the 2×2 probe laser array after laser beams passing through adjacent grating splitters with high diffraction efficiency (99.31%) and effective laser intensity uniformity (90.41%). The laser-MW-coupled atom is described by a four-level model, as illustrated in Fig. 2(d). The probe laser of frequency $\omega_P$ resonantly drives the atomic transition from an atomic ground state $6S_{1/2}$, $F=4$ to an intermediate state $6P_{3/2}$, $F'=5$ with Rabi frequency $\Omega_P$. It has Rabi frequencies of $2\pi \times 4.96$ MHz and a power of 30 μm. A 135-mW strong coupling laser of frequency $\omega_C$ resonantly drive the atomic transition from $6P_{3/2}$, $F'=5$ to a Rydberg state $44D_{5/2}$ for an EIT configuration with Rabi frequency $\Omega_C$ ($2\pi \times 0.56$ MHz Rabi frequencies). The MW field with frequency $\omega_L$ couples two Rydberg states $44D_{5/2}$ and $45P_{3/2}$ with Rabi frequency $\Omega_L$.

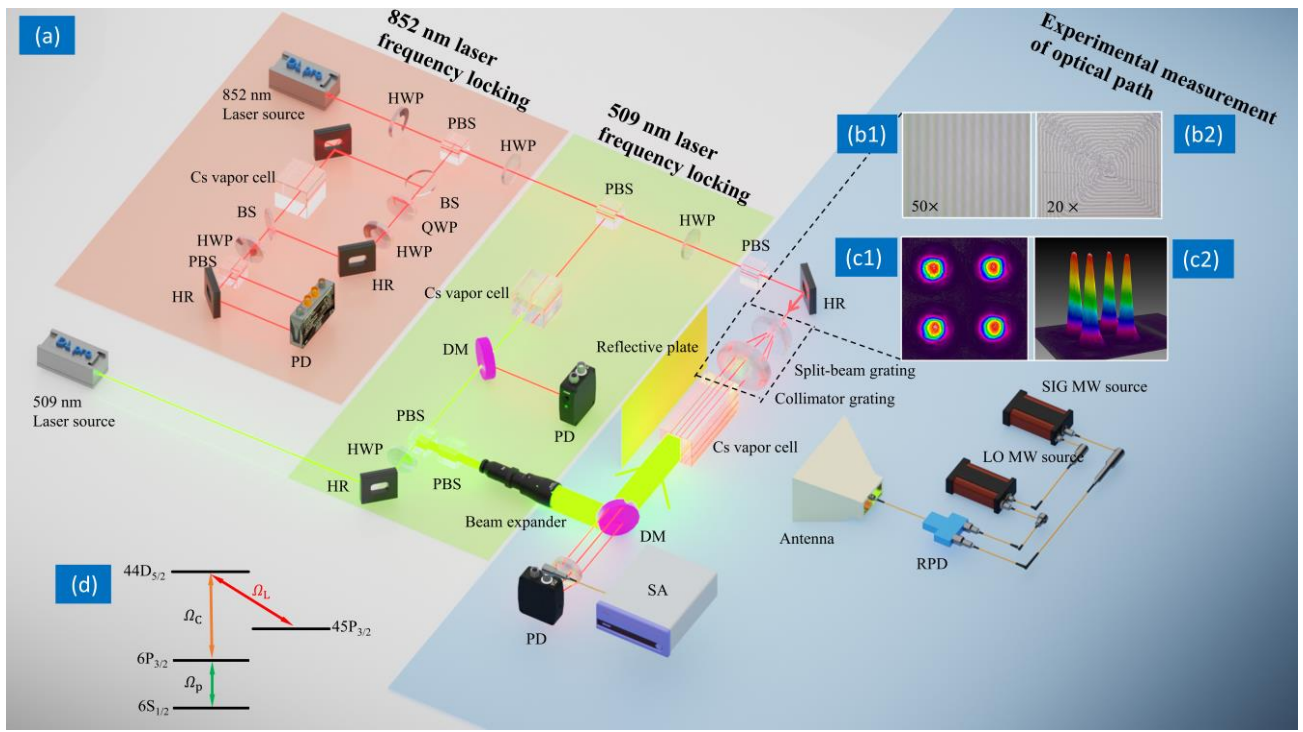

Fig. 2 | (a) Overview of the experimental setup. We have used the following notations: DL PRO: external-cavity diode lasers, HWP: half-wave plate, PBS: polarizing beam splitter, QWP: quarter-wave plate, BS: beam splitter, HR: dielectric mirror, DM: dichroic mirror, SA: spectrum analyzer (Keysight N9020B), RPD: 2-way microwave resistive power divider, SIG: a weak signal generator, LO: local oscillator, BD: balanced detector(Thorlabs PDB210A), PD: photodetector (Thorlabs PDA36A2), (b1, b2) Optical microscope image of the split-beam grating(50 times magnification) and collimator grating(20 times magnification), (c1, c2) beam profile of the 2×2 probe laser array. (d) Experimental energy scheme. An 852 nm probe laser excites the cesium atoms from the ground state $6S_{1/2}$ to the intermediate state $6P_{3/2}$, a 509 nm coupling laser drives the atoms from the intermediate state to the Rydberg state $44D_{5/2}$, and the 8.57 GHz microwave excites the atoms from the Rydberg state $44D_{5/2}$ to the Rydberg state $45P_{3/2}$.

**Results**

Fig. 3(a) illustrates two distinct peaks correspond to the transitions from $6S_{1/2}$ to two allowed $6P_{3/2}$ fine-structure levels ($44D_{3/2}$ and $44D_{5/2}$). Additionally, it demonstrates the rise evolutions of EIT signals height with the increment of probe laser beam numbers illuminating on a PD, due to the growth of Rydberg atom populations. The EIT signals height rises from 0.055 V (25%) in one probe beam to 0.220 V (100%) in four probe beams. This implies that the responsivity is maximum (maximum slope amplitude) at coupling laser resonance ($\Delta_c = 0$) for four beams.

According to the time-domain waveform, the probe laser transmission signals are direct summation of photocurrents with $\delta_S$ = 10 kHz beat-note frequency illuminating a PD, as shown in Fig. 3(b). This experimental finding corresponds with the physical process described by Eq. 2. When multiple same probe laser beams are illuminated a PD, the photocurrent voltage amplitude rises linearly with increment of the beams number.

In order to improve the E-field measurement SNR when weak SIG signal power is fixed, we optimize beat-note frequency to 100 kHz. [24, 42] The SA's high low-frequency 1/f noise led to the selection of 100 kHz following optimization in this experiment. [22] When the intermediate frequency signal has a low frequency, the oscilloscope can be set to receive it.

Furthermore, the *SNR* enhancement distinction is investigated between a 2×2 probe laser array and single-beam laser. The total power grows by 12 dB when a beam is changed to a 2×2 probe laser array. The probe laser optical read-out noises are significantly greater than the PD amplifier noise and the spectrum analyzer noise (by at least 15 dB). Exciting more ground-state atoms to transition into Rydberg states is advantageous for maximizing responsivity. We have optimized the power of the single-beam laser. When the power of the single laser exceeds the optimal level, the Rydberg number remains constant with further increases in power, owing to the Rydberg blockade effect. [46,47]

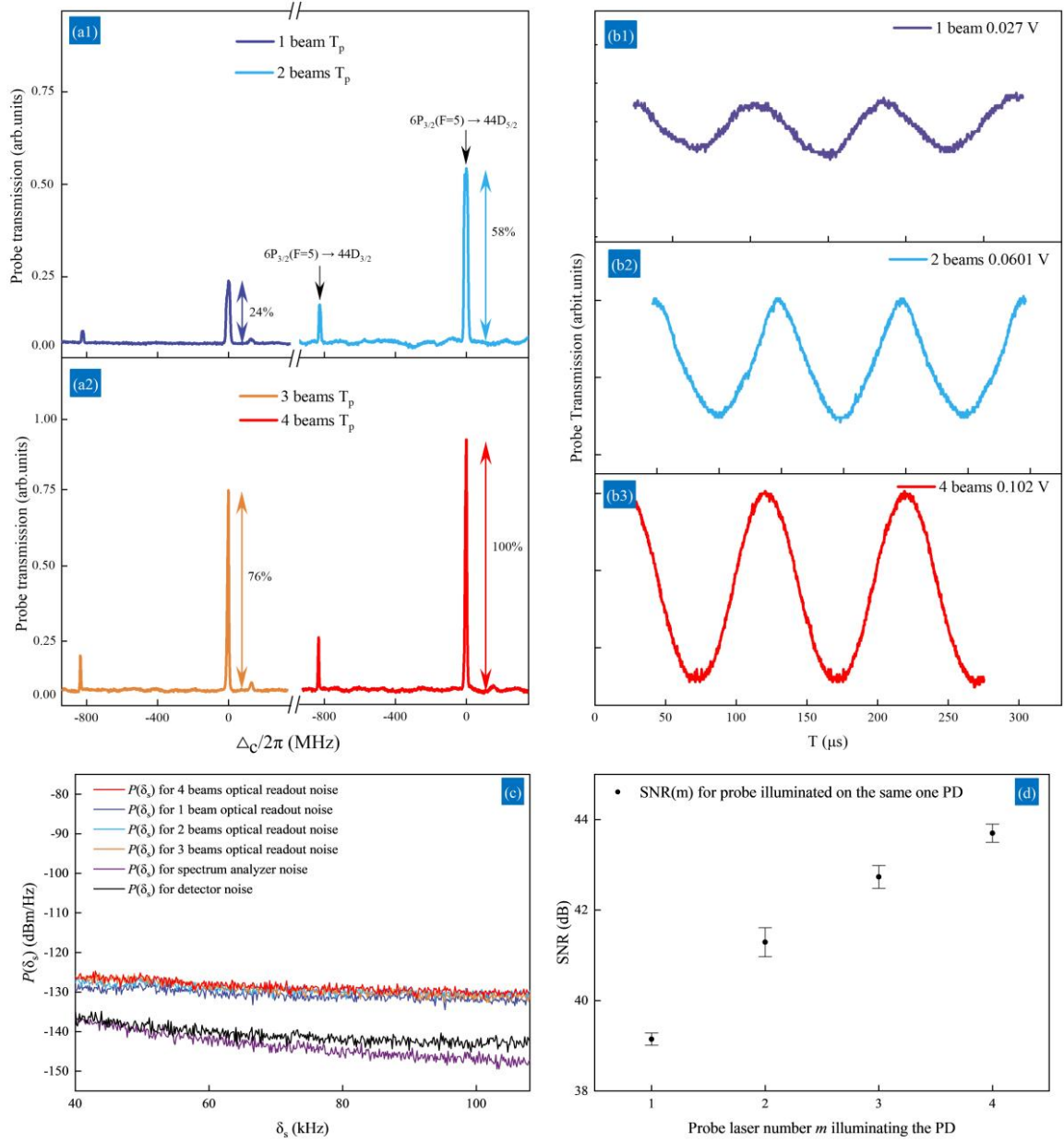

Fig. 3 | (a) Probe laser transmission as a function of coupling laser detuning $\Delta_c$ at different beam numbers. (b) The probe laser transmission signals of $\delta_S = 10$ kHz beat-note frequency at the time domain for 1, 2 beams. (c) The noise spectrum of the Rydberg-atom microwave receiver. Measured NPS as a function of relative signal frequency $\delta_S$ is shown for various number of beams optical readout noises (deep blue, sky blue, brown, red curve), one laser noise with the power as the 2×2 probe laser array (green curve), the spectrum analyzer noise (purple curve) and the amplifier noise of PD (black curve). (d) The measured *SNR* versus probe laser number *m* illuminating the one PD, theoretical curve (red dotted line) and conventional a probe laser illuminating the one PD (gray dotted line).

However, the noise floor rises rapidly, leading to a decrease in the *SNR*. In a laser array, the number of beams grows from one to four, causing a total noise increase of 2.1 dB at 100 kHz. As illustrated in Fig. 3(c), the dark blue represents a single beam laser optical readout noise at -131.7dBm at 100 kHz, whereas the red curve corresponds to a 4-beam laser optical readout noise at -129.6dBm at 100 kHz. This is because the MW E-field inhomogeneity within the vapor cell do not fluctuate with the Rydberg atoms populations, as well as the other types of noise except for the interaction noise. In essence, employing a laser with a high-power density increases the number of photons detected. As a result, more photoelectrons are generated and move around in the PD, increasing the photocurrent noise level. Furthermore, the PD temperature also increases, causing a more active migration of electrons as the power density increases. This demonstrates that a 2×2 probe laser array has lower optical read-out noise than a single beam due to its lower power density within the array under the same total power of a laser beam and the 2×2 laser array illuminating a PD.

To validate the theoretical correctness of employing a laser array to illuminate a single PD for enhancing the *SNR*, we compared the *SNR* as a function of the number of probe lasers (m) illuminating a single PD in Fig. 3(d). This experiment illustrates that the *SNR* of the Rydberg-atom microwave receiver rises with the increment of probe laser numbers. The *SNR* grows 4.6 dB from a single probe laser to a 2×2 laser array.

A copper reflecting plate is utilized to demonstrate that the laser array technique can significantly improve the E-field measurements sensitivity even after the E-field local enhancement has reached its maximal enhancement capabilities. This metal reflector forms a standing wave system, which results in theoretically a twofold local amplification of the E-field. To achieve peak sensitivity, the beat-note frequency $\delta_s$ and the strength of the LO E-fields $E_{LO}$ need to be optimized by fixing the detected SIG signal strength and tuning the $\delta_s$ or $E_{LO}$ for maximum optical output. In our demonstration, the optimal $\delta_s$ is 100 kHz and $E_{LO}$ needs to be adjusted depending on specific experimental conditions. Optimal LO is obtained to maximize responsivity (maximize the probe transmission vs $\Omega_{MW}$ slope) through optimization. Initially, we evaluate numerous EIT-AT splitting spectrum signals of probe transmission generated by distinct LO field in order to quantitatively connect the E-field with spectral splitting in the frequency domain. Fig. 4(a) illustrates EIT signals as a function of the coupling laser detuning. The original EIT peak splits into two transparent absorption peaks. The optimal LO E-field intensity is set to 0.53 V/m (8.58 MHz splitting peaks) at the black curve position in Fig.4 (a).

To precisely compute the E-field for weaker SIG microwave power, adopt the extrapolation via a linear relationship where the E-field is proportional to the square root of the output power, as described by Eq. 4. The linear coefficient

$k_a = 0.39595$ V/m/mW$^{1/2}$ can be calculated via linear fitting, which is plotted in red solid line in Fig. 4(b) Therefore, the weak SIG power $P$ can be directly converted to the E-fields amplitudes.

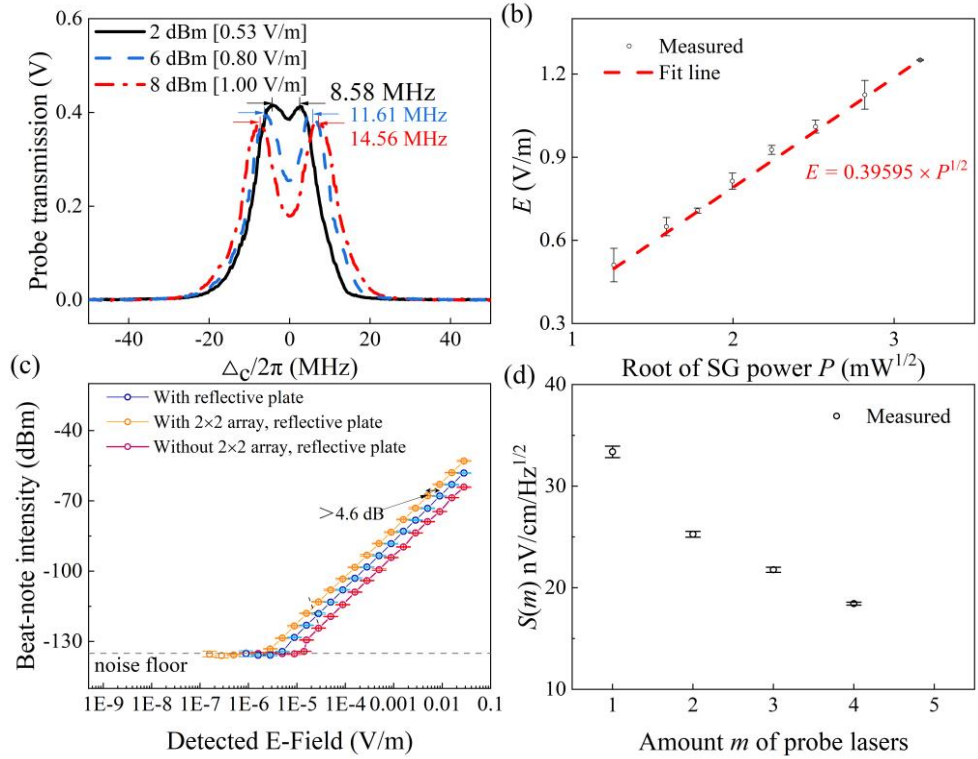

Fig. 4 | (a) Typical experimental data for EIT-AT splitting curves versus coupling laser detuning. (b) Linear relationship between $E$ and $P^{1/2}$ and the error bars represent the standard deviation of $E$. (c) The 100 kHz beat-note intensity of the Rydberg atomic microwave receiver as a function of the incident detected E-fields and the error bars represent the standard deviation of beat-note signal intensity. The spectrum analyzer output signal showed a linear relationship with the incident detected E-fields. (d) The measured sensitivity versus probe laser numbers $m$ and the error bars represent the standard deviation of $S$.

Fig. 4(c) depicts the beat-note signal power measured by the spectrum analyzer is recorded as a function of detected SIG E-fields with and without copper reflective plate and 2×2 probe laser array. The received beat-note signal power is approximately linear proportional to the detected SIG E-fields. The minimum detected SIG E-fields are obtained as sensitivity($S$) by identifying the intersections of the beat-note power curves and the spectrum analyzer noise floor. Furthermore, $S$ were measured to be 65 nV /cm/Hz$^{1/2}$, 33 nV /cm/Hz$^{1/2}$ (with copper reflective plate) and 19 nV /cm/Hz$^{1/2}$(with copper reflective plate and the 2×2 probe laser array), respectively.

To obtain the specific $S$ performance comparison with different probe laser numbers, Fig. 4(d) illustrates that increasing the number of probe lasers results in a higher $S$. The $SNR$ enhances by approximately 5 dB when the number of probe lasers is doubled and illuminating on a single PD. Consequently, these good agreements with theoretical predictions validates that the proposed method remains effective when the microwave enhancement method of increasing sensitivity reaches its limitation. This indicates that if the sensitivity can be raised by 100 times through a split ring resonator, [29] it could be further enhanced to 300 times with the implementation of a 2×2 probe laser array and a split ring resonator. However, achieving 300 times sensitivity enhancement is challenging solely through a resonator.

Table 2 Performance comparisons of Rydberg atomic receiver

| References | Atomic species | Fre. [GHz] | $\mu$ [$ae_o$] | $L$ [cm] | Res. eha. factor | $m$ | $S$ [nV/cm/Hz$^{1/2}$] |
|---|---|---|---|---|---|---|---|
| [40] | Rb | 14 | 1774 | 10 | no eha. | 1 | 8330 |
| [20] | Cs | 5.05 | 1786 | 4 | no eha. | 1 | 5000 |
| [41] | Cs | 5.05 | 1786 | 3 | no eha. | 1 | 3000 |
| [32] | Rb | 16.6 | 1565 | 10 | no eha. | 1 | 49 |
| [22] | Cs | 9.93 | 1137 | 5 | no eha. | 1 | 55 |
| [29] | Cs | 1.3 | 4375 | 1 | 100 | 1 | 55 |
| [42] | Cs | 0.946 | | 2 | 40 | 1 | 213 |
| [43] | Cs | 5 | | 5 | no eha. | 1 | 712 |
| Our work | Cs | 8.57 | 1255 | 5 | 1.9 | 4 | 19 |

We have used the following notations: Fre.: frequency and Res. eha. factor: resonator enhancement factor. $\mu$ is the transition dipole moment, $L$ is the length of vapor cell, $m$ is the probe laser number and $S$ is sensitivity.

To demonstrate the novelty and advantages of this work, Table 2 summarizes the specific performance comparison with Rydberg atomic receiver in previous literatures including measurement frequency, transition dipole moment, the vapor cell length, resonator enhancement factor, the probe laser number and sensitivity. It is shown that, the $S$ of our work is significantly improved compared with conventional Rydberg atomic receivers for same atomic species (Cs), [20, 22, 41] even under high resonator enhancement factor. [29, 42] Moreover, the sensitivity can be further improved when we select Rydberg states with high the transition dipole moment. [24, 43] This implies that the sensitivity would be higher if the present experiment selects the Rydberg state with a higher transition dipole moment but the aspect of the method lies outside the scope of this work. In brief, our work exhibits the high sensitivity among the Rydberg atomic receiver under consideration.

A fair comparison of atomic and conventional microwave receivers can be achieved through establishing a sensitivity conversion relation. In this paper, we use the E-field intensity as a benchmark. The -174 dBm/Hz sensitivity of conventional receivers and gain of 1.5 [45] correspond to 1 nV/cm/Hz$^{1/2}$@ 8.57 GHz (Supplementary Information S2 for the derivation). This is almost five times more sensitive than the most advanced Rydberg receiver in research. [44] Therefore, Rydberg-atom receivers require $SNR$ enhancement to compete with conventional receivers. [45] If the 2×2 probe laser array and reflective plate enhancement are applied to the most advanced sensitivity, [44] the sensitivity is improved from 5.1 nV/cm/Hz$^{1/2}$ to 0.89 nV/cm/Hz$^{1/2}$. This indicates that atomic microwave receivers are more sensitive than conventional receivers. Looking ahead, the *SNR* of atomic receivers can be further enhanced by employing more laser beams and incorporating resonators with higher enhancement factor.

## Conclusions

In summary, a laser array-based approach has been reported to improve the *SNR* of the atomic receiver for the first time. To start with, *SNR* is improved 4.6 dB utilizing more beams (the 2×2 probe laser array) illuminating on a PD. Furthermore, the E-field measurement sensitivity reaches 19 nV/cm/Hz$^{1/2}$@8.57 GHz for the 44D$_{5/2}$ state when we utilize the laser array and the microwave domain approach, which include a microwave LO and a copper reflective plate for microwave E-field local enhancement. In reality, more sensitive atomic receivers may have higher sensitivity. Specifically, sensitivity will be improved from 5.1 nV/cm/Hz$^{1/2}$ to 0.89 nV/cm/Hz$^{1/2}$ adding a 2×2 probe laser array and a reflective plate enhancement into existing the most advanced Rydberg sensitivity [44]. It means that atomic microwave receivers will now surpass conventional receivers in terms of sensitivity. Thus, future iterations of utilization more number laser beams will make atomic receivers more sensitive integrating resonators with higher enhancement factor, combining muti-photon, [20] doppler free [21] and repump [33]. Each probe laser is coherent in array. Hence, the SNR can be enhanced while detecting a weak target signal in strong noise and clutter environment. In the future, this method's ability to prompt SNR will be investigated in external noise-dominated scenarios. Therefore, this work is vital to serve as an excellent guidance for the design and optimization of the ultrahigh-sensitivity Rydberg atomic microwave receiver.

## Acknowledgements

We are grateful for financial supports from National Natural Science Foundation of China (Grant Nos. 12304436, 12104509, 62105338, 12074433) and Natural Science Foundation of Hunan Province of China (Grants No. 2023JJ30626).

## Author contributions

Yunqi Fu proposed the original idea and supervised the project. Bo Wu and Qiang An fabricated the samples and performed the measurements. Di Sang, Ruiqi Mao, Zhanshan Sun, Yi Liu, Yanli Zhou, Yi Lin, and Yunqi Fu gave some important suggestions about the calculation methods. All authors reviewed and approved the final manuscript.

Thanks for the simulation process DongJun Technology Co., Ltd. Eastwave Software provides support and assistance.

## Competing interests

The authors declare no competing financial interests.